\documentclass[12pt, prd, showpacs]{revtex4}

\usepackage{amsmath}
\usepackage{amssymb}
\usepackage{graphicx}
\usepackage{bm} 

\begin{document}

\title{Black hole mimickers: regular versus singular behavior}
\author{Jos\'{e} P. S. Lemos}
\affiliation{Centro Multidisciplinar de Astrof\'{\i}sica, CENTRA,
Departamento de F\'{\i}sica, 
Instituto Superior T\'ecnico - IST, Universidade T\'{e}cnica de Lisboa
- UTL, Avenida Rovisco Pais 1, 1049-001 Lisboa, Portugal\,\,}
\email{lemos@fisica.ist.utl.pt}
\author{Oleg B. Zaslavskii}
\affiliation{Astronomical Institute of Kharkov V. N. Karazin National
University, 35
Sumskaya St., Kharkov, 61022, Ukraine}
\email{ozaslav@kharkov.ua}

\begin{abstract}
Black hole mimickers are possible alternatives to black holes, they
would look observationally almost like black holes but would have no
horizon. The properties in the near-horizon region where gravity is
strong can be quite different for both type of objects, but at
infinity it could be difficult to discern black holes from their
mimickers. To disentangle this possible confusion, we examine the
near-horizon properties, and their connection with far away asymptotic
properties, of some candidates to black mimickers. We study
spherically symmetric uncharged or charged but non-extremal objects,
as well as spherically symmetric charged extremal objects. Within the
uncharged or charged but non-extremal black hole mimickers, we study
non-extremal $\varepsilon$-wormholes on the threshold of the formation
of an event horizon, of which a subclass are called black foils, and
gravastars.  Within the charged extremal black hole mimickers we study
extremal $ \varepsilon$-wormholes on the threshold of the formation of
an event horizon, quasi-black holes, and wormholes on the basis of
quasi-black holes from Bonnor stars. We elucidate, whether or not the
objects belonging to these two classes remain regular in the
near-horizon limit. The requirement of full regularity, i.e., finite
curvature and absence of naked behavior, up to an arbitrary
neighborhood of the gravitational radius of the object enables one to
rule out potential mimickers in most of the cases. A list ranking the
best black hole mimickers up to the worse, both non-extremal and
extremal, is as follows: wormholes on the basis of extremal black
holes or on the basis of quasi-black holes, quasi-black holes,
wormholes on the basis of non-extremal black holes (black foils), and
gravastars. Since, in observational astrophysics it is difficult to
find extremal configurations (the best mimickers in the ranking),
whereas non-extremal configurations are really bad mimickers, the task
of distinguishing black holes from their mimickers seems to be less
difficult than one could think of it.
\end{abstract}

\keywords{quasi-black holes, black holes, wormholes one two three}
\pacs{04.70Bw, 04.20.Gz}
\maketitle



\newpage

\section{Introduction}

In recent years, it has been debated in the literature about possible
alternatives to black holes, the black hole mimickers, which would
look observationally almost like black holes but would have no
horizon. The existence of such objects can, in principle, put in doubt
astrophysical data which otherwise are considered as observational
confirmation in favor of black holes \cite{abr}. On one hand, it is
clear that the properties in the near-horizon region where gravity is
strong can be quite different for both type of objects. On the other
hand, the statements about the difficulties in discerning black holes
from their mimickers are usually related to measurements at spatial
infinity. Thus, one should insist on the question: Can an observer at
infinity catch the difference between both type of objects in some
indirect way, or even rule out some possible mimicker? In our view,
the answer is positive and is connected with key properties, namely,
regularity or singularity, of the corresponding geometries. It turns
out that the requirement of full regularity up to an arbitrary
neighborhood of the gravitational radius of the object enables one to
rule out the potential mimickers in most of the cases.

The goal of the present work is to examine the near-horizon
properties, and their connection with far away asymptotic properties,
of some candidates to black mimickers. We study spherically symmetric
configurations, and make, two major divisions, or classes, on those
candidates. First, uncharged or charged but non-extremal objects, and
second extremal objects. Within the uncharged or charged but
non-extremal one can invoke as black hole mimickers, non-extremal
$\varepsilon$-wormholes on the threshold of the formation of an event
horizon, some of which are called black foils \cite{ds} (see \cite{sm}
for the construction with other purposes of $\varepsilon$-wormholes,
which actually can also act as mimickers), and gravastars \cite{grav}. 
Within the extremal charged class one can invoke extremal $
\varepsilon$-wormholes on the threshold of the formation of an event
horizon, quasi-black holes \cite{qbh} (see also
\cite{lemoszanchinjmp}), and wormholes on the basis of quasi-black
holes from Bonnor stars, to name a few. We want to elucidate, whether
or not the objects belonging to these two classes remain regular in
the near-horizon limit. The arguments of \cite{qbh} which rule out
non-extremal limiting configurations as becoming singular do not apply
to the wormhole case \cite{ds}. Thus, we carry out the corresponding
analysis anew for both classes of objects.

\section{Equations and setup for mimickers}

For our purposes we write a generic spherically symmetric metric as
\begin{equation}
ds^{2}=-\exp \,\left( 2\Phi (r,\lambda _{i})\right) \,dt^{2}+\frac{dr^{2}}{
V(r,\lambda _{i})}+r^{2}(d\theta ^{2}+\sin ^{2}\theta d\phi ^{2})\,,
\label{1}
\end{equation}
where $r$ is the radial coordinate, and the $\lambda _{i}$ are generic
parameters, that depend on each situation we are studying. For instance when
treating wormholes one has $i=1,...,4$, and $\lambda
_{i}=(r_{+},r_{-},\varepsilon ,r_{0})$, such that $\exp (2\Phi (r,\lambda
_{i}))=\exp (2\Phi (r,r_{+},r_{-},\varepsilon ,r_{0}))$ and $V(r,\lambda
_{i})=V(r,r_{+},r_{-},\varepsilon ,r_{0})$. Here $r_{+}$ is the radius of
the would be horizon when $\varepsilon $ is zero, $r_{-}$ is the radius of
the other possible horizon, $\varepsilon $ is in principle a small quantity,
and $r_{0}$ is the radius of a possible matter shell, satisfying $r_{0}\geq
r_{+}$. In other situations, e.g., for gravastars, one has $\lambda
_{i}=r_{0}$ and, possibly, the energy density and the pressure should be
somehow included. For the metric (\ref{1}) the components of the Riemann
tensor in an orthonormal frame, the hat frame, are equal to,
\begin{equation}
K(r)\equiv R_{\hat{t}\hat{r}}^{\hat{t}\hat{r}}=-V(\Phi ^{\prime \prime
}+\Phi ^{\prime 2})-\frac{V^{\prime }}{2}\Phi ^{\prime }\,,  \label{K(r)}
\end{equation}
\begin{equation}
N(r)\equiv R_{\hat{t}\hat{\theta}}^{\hat{t}\hat{\theta}}=-\frac{V}{r}\Phi
^{\prime }\,,  \label{N(r)}
\end{equation}
\begin{equation}
F(r)\equiv
R_{\hat{\phi}\hat{\theta}}^{\hat{\phi}\hat{\theta}}=\frac{1}{r^{2}
}(1-V)\,,  \label{F(r)}
\end{equation}
\begin{equation}
H(r)\equiv
R_{\hat{\theta}\hat{r}}^{\hat{\theta}\hat{r}}=-\frac{V^{\prime }}{
2r}\,\text{,}  \label{H(r)}
\end{equation}
where a prime denotes derivative with respect to $r$. Here these
components of the Riemann tensor have a simple physical meaning. The
$K(r)$ component in Eq. (\ref{K(r)}) yields the radial geodesic
deviation, the $N(r)$ component in Eq. (\ref{N(r)}) yields the angular
deviation, and analogously for the $F(r)$ and $H(r)$ components in
Eqs. (\ref{F(r)}) and (\ref{H(r)}).  In summary, they describe the
deviation of geodesics in the corresponding directions. In general,
forcing a matching at $r_{0}$, with $r_{0}>r_{+}$, surface stresses
$S_{a}^{b}$ appear, which, in a coordinate frame, are equal to
\cite{isr,MTW}
\begin{equation}
8\pi \Sigma \equiv -8\pi S_{t}^{t}=-\frac{2}{r_{0}}\left[ \left(
\frac{dr}{dl
}\right) _{+}-\left( \frac{dr}{dl}\right) _{-}\right] \text{,}  \label{00}
\end{equation}
\begin{equation}
8\pi S\equiv 8\pi S_{\theta }^{\theta }=\frac{1}{r_{0}}\left[ \left(
\frac{dr
}{dl}\right) _{+}-\left( \frac{dr}{dl}\right) _{-}\right] +\left( \frac{
d\Phi }{dl}\right) _{+}-\left( \frac{d\Phi }{dl}\right) _{-}\,,  \label{22}
\end{equation}
$S_{\phi }^{\phi }$ being equal to $S_{\theta }^{\theta }$, and $l$ being
the proper radial distance.
Now, if metric (\ref{1}) represents a wormhole, than the areal radius $r(l)$
should have a local minimum at the throat. Thus, we have two branches 
emerging out of the minimum radius, one with $\left(
\frac{dr}{dl}\right) _{+}=\sqrt{V(r)}$ and 
the other with $\left( \frac{dr}{dl}\right)
_{-}=-\sqrt{V(r)}$. Then,
\begin{equation}
8\pi \Sigma =-\frac{4}{r_{0}}\sqrt{V(r_{0})},  \label{s00}
\end{equation}
\begin{equation}
8\pi S=\frac{2\sqrt{V(r_{0})}}{r_{0}}+\left( \frac{d\Phi }{dl}\right)
_{+}-\left( \frac{d\Phi }{dl}\right) _{-}\,\text{.}  \label{s22}
\end{equation}
There are also the bulk stress-energy components, but those do not
interest us here and do not need computation.

In some situations one has to deal here with naked behavior. This
means there are cases in which the Kretschmann scalar and other
curvature quantities are finite on the horizon in a static coordinate
system, but some of those quantities may blow up in a freely falling
frame. Such a kind of behavior is called naked behavior, and many
instances of it have been found 
\cite{nk1,nk2,br1,boothmann,vo,tr,sm2,odintsov}. Another
example is with quasi-black holes \cite{qbh}. One of the features
typical of quasi-black holes, consists in precisely showing naked
behavior on and beyond the quasihorizon surface. In addition, metrics
obtained by gluing two spacetimes, can have a similar behavior, but
now the surface stresses, which are finite in a static coordinate
frame, blow up in a free-falling frame. Thus, since we have found in
Eqs. (\ref{K(r)})-(\ref{22}) the curvature and surface stresses in a
static frame for the spacetime in question, we now examine the
behavior of the same quantities for a free-falling frame. Consider
then a radial local boost from a static frame with four-velocity
$u^{\mu }$ to a free-falling frame with the velocity $ \bar{u}^{\mu
}$. Under a boost the four-velocity transforms according to $
\bar{u}^{\mu }=u^{\mu }\cosh \alpha -n^{\mu }\sinh \alpha $, where the
orthonormal vector $n^{\mu }$ is pointing in the radial outward
direction, and $\alpha $ is the velocity boost parameter. In relation
to the tidal forces in the bulk, the curvature components
(\ref{K(r)})-(\ref{N(r)}) in the orthonormal basis responsible for
tidal forces transform according to
\begin{equation}
\bar{K}=K\,,
\end{equation}
\begin{equation}
\bar{N}=N-Z\,\sinh ^{2}\alpha =H+E^{2}\exp (-2\Phi )(N-H)\,,  \label{barN}
\end{equation}
where a bar means a quantity evaluated in the freely falling frame, $Z=H-N$
(see equation (\ref{H(r)}) for the definition of $H$), $\cosh \alpha =\exp
(-\Phi )\,E$, and $E$ is the energy of the particle frame (see e.g. \cite
{qbh} for more details). The most interesting situation arises when $K$ is
finite (so, the Kretschmann scalar is also finite) but $\bar{N}$ diverges.
The corresponding horizons can be called truly naked \cite{vo,tr}. In
relation to the surface stresses, it is useful to define the quantity
\begin{equation}
\bar{\Sigma}=S_{\mu \nu }\bar{u}^{\mu }\bar{u}^{\nu }\,, 
\end{equation}
which represents the energy density of the shell as observed by the observer
with the four-velocity $\bar{u}^{\mu }$. In a static frame $\bar{\Sigma}
=\Sigma =-S_{t}^{t}$. Then, considering a boosted motion along a radial
geodesic with energy $E$, one obtains
\begin{equation}
\bar{\Sigma}=-S_{t}^{t}\,\exp (-2\Phi )\,E^{2}\,. \label{sigmacapital}
\end{equation}
This is a useful
expression for analyzing naked behavior of wormholes and other objects. For
the wormhole case it reduces to
\begin{equation}
\bar{\Sigma}=-\frac{\sqrt{V(r_{0})}}{2\pi r_{0}}\,\exp (-2\Phi )\,E^{2}\,,
\label{sigmacapital2}
\end{equation}
where $r_{0}$ is the radius at which the shell is located,
and we took into account Eq. (\ref{s00}).

\section{Mimickers of non-extremal black holes}

\subsection{Non-extremal wormholes on the basis of 
$\protect\varepsilon$-metrics with surgery}

\label{nonextremalepsilonsurg}

\subsubsection{Basics}

\noindent 
There are many ways of making wormholes \cite{th,vis}. In this section we
are interested in making wormholes from charged metrics, more general than
the Reissner-Nordstr\"{o}m metric, but for certain choices the metrics can
be reduced to the Reissner-Nordstr\"{o}m metric. Even for this metric one
probably can think of many manners of making wormholes. We are interested in
two different ways, that easily lead to the threshold of black hole
formation, and the discussion of how they mimic black holes. Then we
compound both ways into one single way.

The first way is the surgery approach, see (see in particular Section
15.2.1 of \cite{vis}). Pick up a spherically symmetric metric of the
form $ ds^{2}=-\exp \,(2\Phi
(r,r_{+},r_{-},r_{0}))\,dt^{2}+\frac{dr^{2}}{
V(r,r_{+},r_{-},r_{0})}+r^{2}(d\theta ^{2}+\sin ^{2}\theta d\phi
^{2})$, where $r_{+}$ is the radius of the would be horizon, $r_{-}$
is the radius of the other possible horizon, and $r_{0}$ is the radius
of a possible matter shell, satisfying $r_{0}\geq r_{+}$. Take for
instance the non-extremal Reissner-Nordstr\"{o}m metric, where $\exp
(\Phi )=\sqrt{V}= \sqrt{(1-\frac{r_{+}}{r})(1-\frac{r_{-}}{r}})$, with
$r_{\pm }=GM\pm \sqrt{ G^{2}M^{2}-G\,Q^{2}}$ and $r_{+}\neq r_{-}$,
$M$ and $Q$ being the mass and electrical charge of the object,
respectively, and $G$ is Newton's constant (we use $c=1$). Cut the
metric at some $r_{0}$ and join the resulting spacetime with a
symmetric branch. This is a non-extremal Reissner-Nordstr\"{o}m
surgery (the Schwarzschild surgery, with no charge and so $r_{-}=0$,
being a particular case of this), resulting in a non-extremal wormhole
with a thin shell of matter at $r_{0}$, the throat. In brief, one
places at some radius, $r_{0}$, a thin shell which separates two
regions, with non-extremal geometries, with $r_{0}$ also defining the
throat. Then, one introduces another radial coordinate $l$, such that
$r=r(l)$ with $r_{0}=r(0) $, and which covers the whole of the
manifold, $-\infty <l<\infty $. The function $ r(l)$ is monotonically
decreasing for the branch $l<0$, which we call the $ ``-"$ branch, and
monotonically increasing for $l>0$, the \textquotedblleft +"
branch. In general, giving this construction, surface stresses
$S_{a}^{b}$ appear.

A second way, i.e., another approach, to build wormholes, is through
metrics of the type $ds^{2}=-\exp \,(2\Phi
(r,r_+,r_-,\varepsilon))\,dt^{2}+\frac{
dr^{2}}{V(r,r_+,r_-,\varepsilon)} +r^{2}(d\theta ^{2}+\sin ^{2}\theta
d\phi ^{2})$, where $\varepsilon$ is a small quantity, $r_+$ is the
radius of the would be horizon when $\varepsilon$ is zero, and $r_-$
is the radius of the other possible horizon \cite{ds,sm}. Metrics of
this type, depending on the parameter $\varepsilon$, can be
generically call $\varepsilon$-spacetimes, which in special cases can
become wormholes, i.e., $\varepsilon$-wormholes.  In \cite{ds} the
model with metric $ds^{2}=-\left(V+\varepsilon^2\right) \,dt^{2}+
\frac{dr^{2}}{V} +r^{2}(d\theta ^{2}+\sin ^{2}\theta \,d\phi ^{2})$,
was considered, where thus $\exp(\Phi(r,\varepsilon))=\sqrt{
V+\varepsilon^{2}}$, with $V$ being chosen appropriately. In turn, in
\cite {sm} the model with metric $ds^{2}=-\left(\lambda \sqrt{V}
+\varepsilon\right)^2\, dt^{2}+\frac{ dr^{2}}{V} +r^{2}(d\theta
^{2}+\sin ^{2}\theta \,d\phi ^{2})$, was considered, where thus
$\exp\,(\Phi)=\lambda \sqrt{V}+\varepsilon$, with $V$ being chosen
appropriately, and with $ \lambda $ being an additional
parameter. Thus, a generic $\varepsilon$-metric of the type given
above yields a generic spacetime that comprehends the two cited
models, one model studied in \cite{ds}, the other in \cite{sm}.  One
can calculate the Riemann tensor for the $\varepsilon$-metric, and of
course, since spacetime is not empty, there is a smooth
energy-momentum tensor associated to the metric but we do not need to
calculate it here. All these $\varepsilon$-spacetimes are smooth. Now,
take the $\varepsilon$-metric to construct a wormhole, i.e, an
$\varepsilon$-wormhole. Since for the construction we need to impose
some more conditions, in particular on the the potentials $\Phi$ and
$V$ of the $\varepsilon$-metric, let us adopt the following
approach. First, as above, one introduces the radial coordinate $l$,
such that $r=r(l)$ and $-\infty <l<\infty$. The function $ r(l)$ is
monotonically decreasing for the $``-"$ branch, $l<0$, and
monotonically increasing for the $``+"$ branch, $l>0$. Second, the
dependence of the function $\Phi$ on the parameter $\varepsilon$,
$\Phi =\Phi(r,\varepsilon)$, which can be of the type of the models
considered above \cite{ds,sm}, is such that $\exp\left(\Phi(r_{+},0)
\right)=0 $, and the dependence of the function $V$ on the parameter
$\varepsilon$ is also such that $V(r_{+},0)=0$, so in the limit
$\varepsilon\rightarrow0$ the original wormhole configuration indeed
approaches a black hole. Third, if the first derivative
$\frac{dr}{dl}$ is continuous at the throat, we have $
\frac{dr}{dl}=0$ (see \cite{th}). When $V$ does not depend on
$\varepsilon$ at all, the throat is situated on the possible would be
horizon. This is the approach used in \cite{ds,sm} to build a
wormhole. This approach is smooth as long as $\varepsilon\neq0$. As a
particular instance of this approach, one can choose $V$ as being
Reissner-Nordstr\"om, $V\equiv (1-\frac{r_+}{r} )(1-\frac{r_-}{r})$,
as usual. For $r_+\neq r_-$ one has a non-extremal choice for $V$, the
case $r_-=0$, i.e., $Q=0$, yielding the Schwarzschild potential $V$ as
a particular case, the one chosen in \cite{ds,sm}. Such $
\varepsilon$-wormholes have been called foils in \cite{ds}. When $
\varepsilon=0$ we have the full non-extremal Reissner-Nordstr\"{o}m
metric.

So, let us compound both approaches, the surgery approach of \cite{th,vis},
and the $\varepsilon$ approach of \cite{ds,sm}. Write then a generic $
\varepsilon$-metric with surgery as $ds^{2}=-\exp \,(2\Phi
(r,r_{+},r_{-},\varepsilon ,r_{0}))\,dt^{2}+\frac{dr^{2}}{
V(r,r_{+},r_{-},\varepsilon ,r_{0})}+r^{2}(d\theta ^{2}+\sin ^{2}\theta
d\phi ^{2})\,,\label{1-sub}$ i.e., from (\ref{1}) one chooses
\begin{equation}
\exp \,(2\Phi (r,\lambda _{i}))=\exp \,(2\Phi (r,r_{+},r_{-},\varepsilon
,r_{0}))\,\,,\;\;V(r,\lambda _{i})=V(r,r_{+},r_{-},\varepsilon ,r_{0})\,,
\label{epsilonwormpot}
\end{equation}
where $\varepsilon$ is a small quantity, $r_{+}$ is the radius of the
would-be horizon (if $\varepsilon=0$), and $r_{0}$ is the radius of a
possible matter shell, satisfying $r_{0}\geq r_{+}$. Since we are
studying here non-extremal metrics we have, when $\varepsilon =0$,
that $V(r_{+})=0$ and $V^{\prime }(r_{+})\neq 0$.  Essentially, what
we have done is a surgery on $\varepsilon $-metrics, non-extremal
Reissner-Nordstr\"{o}m (with Schwarzschild included) metrics being
particular $\varepsilon =0$ instances. This can also be thought of as
a one-parametric deformation of the original $\varepsilon $-wormhole
metric by gluing two branches at the throat $r_{0}$, with
$r_{0}>r_{+}$. Now, given the general $\varepsilon $-metric with
surgery, (\ref{1}) and (\ref {epsilonwormpot}), and the correspondent
wormhole construction, we are interested in getting a spacetime that
mimics a black hole. It is then not hard to understand that there are
two distinct situations to obtain spacetimes on the threshold of being
black holes. If there is no shell, then wormholes approach black holes
when $\varepsilon \rightarrow 0$. If there is a shell but $\varepsilon
=0$ then the wormhole throat approaches the horizon when
$r_{0}\rightarrow r_{+}$. Therefore, there is a play of two small
parameters $\varepsilon $ and $r_{0}-r_{+}$ and the limiting procedure
should be considered with great care. It gives rise to two distinct
situations, depending on the order one takes the limiting procedures.
Situation BT: This situation is achieved by, in the end turning the
wormhole metric into the metric of a black hole (B) (i.e., taking
$\varepsilon \rightarrow 0$ as the last operation), after first having
moved the shell towards the minimum throat (T) radius (i.e., taking
$r_{0}\rightarrow r_{+}$ as the initial operation). Formally, this
means taking the limits in the following order $\mathrm{BT}\equiv
\lim_{\varepsilon \rightarrow 0}\lim_{r_{0}\rightarrow
r_{+}}$. Situation TB: This situation is achieved by, in the end the
location of the shell approaches the throat (T) (i.e., taking
$r_{0}\rightarrow r_{+}$ as the last operation), after first 
turning the wormhole metric into a black hole (B) (i.e.,
taking $\varepsilon \rightarrow 0$ as the initial
operation). Formally, this means taking the limits in the following
order $\mathrm{TB}\equiv \lim_{r_{0}\rightarrow
r_{+}}\lim_{\varepsilon \rightarrow 0}$.
Note the case considered in \cite{ds} for the metric (\ref{1}) is a
particular instance of the BT situation, since there $r_{0}=r_{+}$ always,
and one only takes the $\varepsilon \rightarrow 0$ limit. So the situation
BT is the one that yields black foils, following the nomenclature of \cite%
{ds}. One can calculate from equations (\ref{N(r)})-(\ref{H(r)}) that the
components $N(r)$, $F(r)$ and $H(r)$ of the Riemann tensor are always
finite, and from equation (\ref{00}) that both limits when applied to $%
\Sigma \equiv -S_{t}^{t}$ give zero, i.e., $\lim_{\varepsilon \rightarrow
0}\lim_{r_{0}\rightarrow r_{+}}\Sigma =0=\lim_{r_{0}\rightarrow
r_{+}}\lim_{\varepsilon \rightarrow 0}\Sigma $. But for the quantities $K$,
$S\equiv S_{\theta }^{\theta }$ and $\bar{\Sigma}$ (see equations
(\ref{K(r)}), 
(\ref{22}) and (\ref{sigmacapital})) the situation may be different
depending on the order one takes the limits.

Now, as we have been seeing, in treating this problem there are many
levels of distinction. First, we can specify two models of
$\varepsilon$-metrics with surgery which depend on the parameter
$\varepsilon$ and $r_{0}$, namely the model considered in \cite{ds},
where $\exp (\Phi )=\sqrt{ V+\varepsilon^{2}}$, and $V$ is
non-extremal, or some appropriate generalization of it, which we will
call Model 1, and the model given in \cite{sm}, where $\exp (\Phi
)=\lambda \sqrt{V}+\varepsilon$ and $V$ non-extremal, or some
appropriate generalization of it, which we will call Model 2. Second,
within each of the two cases provided by Model 1 and Model 2, we
should study the situations BT and TB. Furthermore, as we want to
examine the regularity of the system under discussion, the relevant
quantities which we are going to calculate are the spacetime curvature
components, and the surface stresses which appear on the glued
boundary, i.e., the shell. We will also study the naked behavior of
each case. So, within each situation we have to study the behavior of
the scalars, and in addition the naked behavior. Thus we have eight
distinct cases to analyze.  We consider these eight cases, each in
turn.

\subsubsection{Models}

\noindent Here we consider the one-parametric deformation,
Eqs. (\ref{1}) and (\ref{epsilonwormpot}), such that for
$\varepsilon=0$ our metric represents the gluing of two non-extremal
black holes. Note that in this non-extremal case the function $V$ of
the metric (\ref{1}) does not need to contain the parameter
$\varepsilon$, so we put $V=V(r_{+},r_{0},r)$.

\hskip 7.4cm \textit{\small Model 1}

\noindent Let the metric have the form (\ref{1}) together with (\ref
{epsilonwormpot}). For Model 1 choose the metric potentials as
\begin{equation}
\exp (\Phi )=\sqrt{V+\varepsilon ^{2}}\,,  \label{efv}
\end{equation}
where $V(r)$ can be any function that satisfies $V(r_{+})=0$ and $V^{\prime
}(r_{+})\neq 0$. For instance $V$ can be Reissner-Nordstr\"{o}m, $V(r)\equiv
(1-\frac{r_{+}}{r})(1-\frac{r_{-}}{r})$ with $r_{\pm }=GM\pm \sqrt{
G^{2}M^{2}-G\,Q^{2}}$ and $r_{+}\neq r_{-}$, with the case $r_{-}=0$ being
the Schwarzschild case, the one chosen in \cite{ds}. Let us now work out
generically the general behavior of this model, i.e., how the curvature and
stress-tensor quantities behave, and also work out generically the naked
behavior. Then we apply these behaviors to the two situations BT and TB. In
doing so, we will display the properties of the quantities $K$, $S$, which
characterize regular or singular general behavior, and the properties of the
quantities $\bar{N}$, $\bar{\Sigma}$ which characterize non-naked or naked
behavior. As for understanding the general behavior note that, from
equations (\ref{K(r)}) and (\ref{22}), explicit calculations give
\begin{equation}
K=-\frac{V}{2}\frac{V^{\prime \prime }}{V+\varepsilon
^{2}}-\frac{1}{4}\frac{
\varepsilon ^{2}V^{\prime 2}}{(V+\varepsilon ^{2})^{2}}\text{,}  \label{e}
\end{equation}
and
\begin{equation}
8\pi S_{{}}=\frac{2}{r_{0}}\sqrt{V(r_{0})}+\frac{V^{\prime }(r_{0})\sqrt{
V(r_{0})}}{\varepsilon ^{2}+V(r_{0})}.  \label{22e}
\end{equation}
In relation to naked behavior, note that since from (\ref{N(r)}) one has
$N=-
\frac{V}{r}\Phi ^{\prime }=-\frac{VV^{\prime }}{2r(\varepsilon ^{2}+V)}$,
and from (\ref{H(r)}) one has $H=-\frac{V^{\prime }}{2r}$, one finds $
Z=H-N=-\varepsilon ^{2}\frac{V^{\prime }}{2r(V+\varepsilon ^{2})}$. Thus,
from (\ref{barN})
\begin{equation}
\bar{N}=\frac{V^{\prime }}{2r}\left[ \frac{E^{2}\varepsilon ^{2}}{
(V+\varepsilon ^{2})^{2}}-1\right] \,.  \label{newbarN}
\end{equation}
With this we can now study the situations BT and TB.

\vskip0.4cm \noindent \textit{Situation BT:} As for the general behavior,
one has that for a non-extremal system $V(r_{+})=0$ and $V^{\prime
}(r_{+})\neq 0$. Then, it follows from equation (\ref{e}) that
\begin{equation}
K(r_{+},\varepsilon )=-\frac{V^{\prime 2}(r_{+})}{4\varepsilon ^{2}}\,,
\label{kinf}
\end{equation}
and so,
\begin{equation}
\lim_{\varepsilon \rightarrow 0}\lim_{r_{0}\rightarrow
r_{+}}K(r_{0},\varepsilon )=-\infty .  \label{einf}
\end{equation}
Correspondingly, the Kretschmann scalar $\mathrm{Kr}=R_{\alpha \beta
\gamma \delta }R^{\alpha \beta \gamma \delta }$ also diverges. It was
shown in \cite {ds} for the choice $V=1-\frac{r_{+}}{r}$ that for
$\varepsilon \neq 0$ there exist geodesics which have no analogue for
the Schwarzschild black hole metric. The timelike particles which move
along them oscillate between turning points which are situated at
different sides of the throat. However, the problem is that in the
limit $\varepsilon \rightarrow 0$ these geodesics pass through a
region of a strong gravitational field. This gives rise to tidal
forces in the radial direction which are of order $\varepsilon
^{-2}r_{+}^{-2}$. If $\varepsilon \ $is exponentially small \cite{ds},
the tidal forces are exponentially large. In addition, from equation
(\ref{22e}), it follows that
\begin{equation}
\lim_{\varepsilon \rightarrow 0}\lim_{r_{0}\rightarrow r_{+}}8\pi
S_{{}}(r_{0},\varepsilon )=0\text{.}  \label{0}
\end{equation}
Now let us analyze the naked behavior. Here one has, $\bar{N}\sim
\varepsilon ^{-2}$, $\bar{K}\sim -\varepsilon ^{-2}\rightarrow $. So,
$\bar{N }\rightarrow \infty $, $\bar{K}\rightarrow -\infty $. Thus,
there is infinite contraction in the longitudinal direction and
infinite transversal stretching. Moreover, since $V(r_{+})=0$ and
$\exp \left( \Phi (r_{+},\varepsilon )\right) \neq 0$, we obtain
immediately from (\ref {sigmacapital2}) and (\ref{efv}) that in the
situation BT one has
\begin{equation}
\bar{\Sigma}=0.
\end{equation}

\vskip0.4cm \noindent \textit{Situation TB:} As for the general behavior,
now one finds, $K(r_{0},0)=-\frac{1}{2}V^{\prime \prime }(r_{0})$. So,
\begin{equation}
\lim_{r_{0}\rightarrow r_{+}}\lim_{\varepsilon \rightarrow
0}K(r_{0},\varepsilon )=-\frac{V^{\prime \prime }(r_{+})}{2}\,,  
\label{k1}
\end{equation}
a result equal to that of a black hole. Also, $8\pi
S_{{}}(r_{0},0)=\frac{2}{ r_{0}}\sqrt{V(r_{0})}+\frac{V^{\prime
}(r_{0})}{\sqrt{V(r_{0})}}$, so \begin{equation}
\lim_{r_{0}\rightarrow r_{+}}\lim_{\varepsilon \rightarrow 0}8\pi
S_{{}}(r_{0},\varepsilon )=+\infty \,.  \label{inf} \end{equation}
Note that equation (\ref{inf}) is in agreement with the behavior of
surface stresses of a wormhole obtained by gluing two copies of the
Schwarzschild metric (see equation 15.46 of \cite{vis}). Now let us
analyze the naked behavior. One has, $\bar{N}=N=-\frac{V^{\prime
}(r_{+})}{2r}$, so $\bar{N}$ is finite and negative. Thus, one obtains
finite deformation in both directions. Moreover, it also follows from
(\ref{sigmacapital}) that in the situation TB
\begin{equation}
\bar{\Sigma}\rightarrow -\infty \,,  \label{barsigmaspecific1}
\end{equation}
i.e., $\bar{\Sigma}$ diverges. Thus, a free-falling observer encounters
diverging surface energy density. The same conclusion applies to the flux $
J=S_{\mu \nu }\bar{u}^{\mu }\bar{e}^{\nu }$.

Concluding here Model 1, we can say that there are two nonequivalent limits
but each of them is \textquotedblleft bad" in that in the BT situation the
Kretschmann scalar diverges, whereas in the TB situation it is the surface
stresses that diverge.

\vskip 0.5cm \hskip 7.4cm \textit{\textit{\small Model 2}} \vskip 0.1cm

\noindent Let the metric have the form (\ref{1}) together with (\ref
{epsilonwormpot}). For Model 2 choose the metric potentials as
\begin{equation} \exp (\Phi )=\lambda \sqrt{V}+\varepsilon \,,
\label{Phimodel2} \end{equation} with $\lambda $ and $\varepsilon$
being parameters. In \cite{sm} the model with
$V=\sqrt{1-\frac{r_{+}}{r}}$, was considered, in which case, when
$\varepsilon=0$ one has the Schwarzschild metric, see also
\cite{sm2}. Here, $ V(r)$ can be any function that satisfies
$V(r_{+})=0$ and $V(r_{+})^{\prime }\neq 0$, a typical example being
the Reissner-Nordstr\"{o}m $V$ potential. As for the general behavior,
again after some calculations, we obtain
\begin{equation}
K(r,\varepsilon )=-\frac{\lambda }{2}\frac{\sqrt{V}V^{\prime \prime }}{
\varepsilon +\lambda \sqrt{V}}\,,
\end{equation}
and
\begin{equation}
8\pi S_{{}}=\frac{2}{r_{0}}\sqrt{V(r_{0})}+\frac{\lambda
V^{\prime }(r_{0})}{
\varepsilon +\lambda \sqrt{V(r_{0})}}\,.
\end{equation}
Now let us analyze the naked behavior. Also, again after some
calculations, we obtain $N(r)=-\frac{V}{r}\Phi ^{\prime }=-\frac{\lambda
\sqrt{V}V^{\prime }}{2r(\varepsilon +\lambda \sqrt{V})}$, so it follows from
(\ref{barN}) that
\begin{equation}
\bar{N}=\frac{V^{\prime }}{2r}\left[ \frac{E^{2}\varepsilon }{(\varepsilon
+\lambda \sqrt{V})^{3}}-1\right]  \label{n2}
\end{equation}
With this we can now study the situations BT and TB.

\vskip0.4cm \noindent \textit{Situation BT:} As for the general behavior, in
this situation one finds,
\begin{equation}
\lim_{\varepsilon \rightarrow 0}\lim_{r_{0}\rightarrow
r_{+}}K(r_{0},\varepsilon )=0\,,
\end{equation}
and
\begin{equation}
\lim_{\varepsilon \rightarrow 0}\lim_{r_{0}\rightarrow r_{+}}8\pi
S_{{}}(r_{0},\varepsilon )=\infty \,.
\end{equation}
Now let us analyze the naked behavior. One has $\bar{K}\rightarrow 0$, and
\begin{equation}
\bar{N}\rightarrow -\infty \,.
\end{equation}
So one finds 
no longitudinal deformation, and an infinite transversal stretching.
Taking into account that $V(r_{+})=0$, $\exp [\Phi (r_{+},\varepsilon )]\neq
0$, we obtain immediately from (\ref{sigmacapital2}) that in situation BT
one finds $\bar{\Sigma}=0$.

\vskip0.4cm \noindent \textit{Situation TB:} As for the general behavior in
this situation, one finds,
\begin{equation}
\lim_{r_{0}\rightarrow r_{+}}\lim_{\varepsilon \rightarrow
0}K(r_{0},\varepsilon )=-\frac{V^{\prime \prime }(r_{+})}{2}\,,  \label{k2}
\end{equation}
and
\begin{equation}
\lim_{r_{0}\rightarrow r_{+}}\lim_{\varepsilon \rightarrow 0}8\pi
S_{{}}(r_{0},\varepsilon )=\infty \,.  \label{s2}
\end{equation}
Now let us analyze the naked behavior. One finds, finite transversal
stretching and finite longitudinal contraction ($V^{\prime \prime
}(r_{+})>0$) or stretching ($V^{\prime \prime }(r_{+})<0$). It follows
from (\ref {sigmacapital2}) and (\ref{efv}) that in the situation TB,
\begin{equation}
\bar{\Sigma}\rightarrow \infty \,,
\end{equation}
i.e., it diverges. Thus, a free-falling observer encounters diverging
surface energy density in the situation TB. The same conclusion applies to
the flux $J=S_{\mu \nu }\bar{u}^{\mu }\bar{e}^{\nu }$.

Concluding here Model 2, we can say that in both situations, BT and TB, the
curvature components remain finite but the limiting values of $K$ do not
coincide. Moreover, in both situations the surface stresses diverge.

\vskip 0.5cm \centerline{\it\small Overall conclusions for Models 1 and 2} 
\vskip 0.1cm

\begin{figure}[h]
\includegraphics*
[width=16.0cm]
{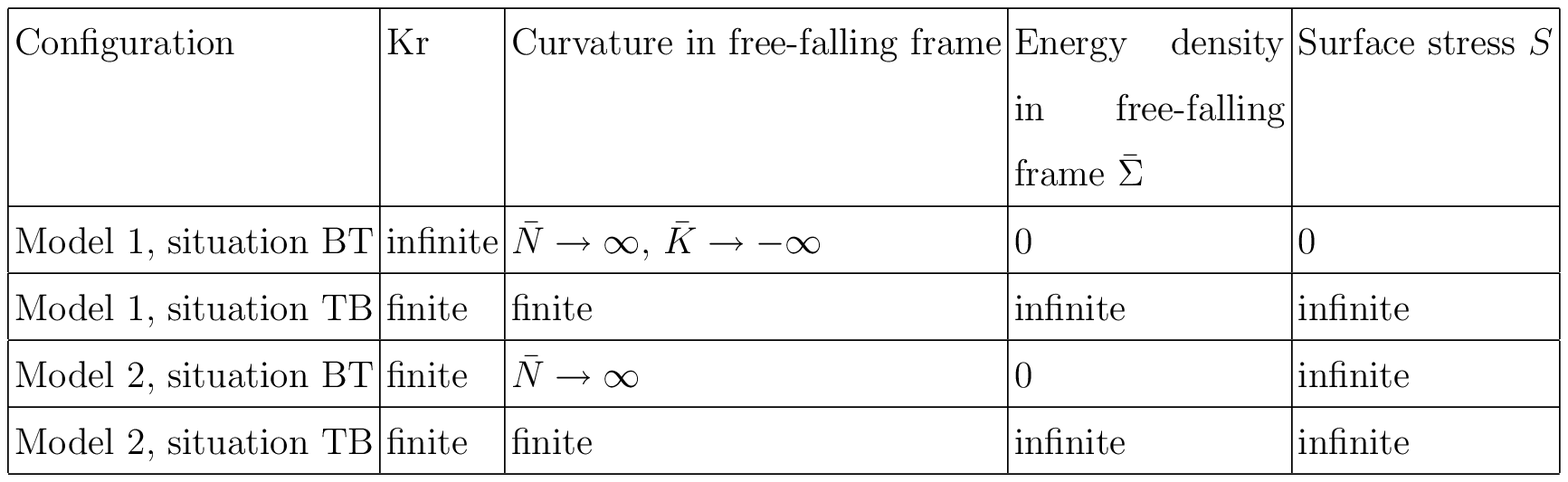}
\caption{\label{table}\small Table summarizing the main features of
the Models 1 and 2 for nonextreme mimickers, in each situation BT or TB, 
studied in the text.}
\end{figure}

\noindent 
As an overall conclusion for the situation TB in Models 1 and 2, we
find the results agree for both models. This can be explained and
generalized as follows. Since, in the situation TB, the limit
$\varepsilon\rightarrow0$ is taken first, the dependence of the metric
on $\varepsilon$ drops out in the final expressions for the curvature
and surface stresses, so any model gives the same result.  In
addition, assuming that for $\varepsilon=0$ one has $\exp (2\Phi )=V$,
and taking into account that $V(r_{+})=0$ and $\exp \left( \Phi
(r_{+},\varepsilon )\right) \neq 0$, we obtain immediately from (\ref
{sigmacapital2}) that in the situation BT, $\bar{\Sigma}=0$, and in
the situation TB, $\bar{\Sigma}$ diverges. This holds independently of
the kind of the model used. Thus, a free-falling observer encounters
diverging surface energy density in the situation TB. The same
conclusion applies to the flux $J=S_{\mu \nu }\bar{u}^{\mu
}\bar{n}^{\nu }$.  In the non-extremal cases just studied, it turned
out that each of the limits under discussion is singular: either the
Kretschmann scalar or surface stresses on the throat (or both)
diverge. Thus, the limit is singular. In other words, a black hole
mimicker made from a wormhole, and in particular a black hole foil, is
not smooth. It is convenient to summarize the results in a table shown
in Figure \ref{table}.
\vskip 3.7cm

\vskip0.5cm
\subsubsection{Remarks: naked behavior and observable differences between
non-extremal black holes and non-extremal $\protect\varepsilon$-wormholes}

\noindent 
Although the singular or almost singular behavior of these black hole
mimickers based on $\varepsilon$-wormholes casts doubts on their real
existence, it is worth our while to study a little more on the effects
of such mimickers on infalling sources and their detection by far away
observers. Intuitively it is clear that there should be some
observational effects if infalling sources are distorted by stronger
than normal tidal fields.

First, we point out that indeed the fact that tidal forces grow
unbound when $\varepsilon \rightarrow 0$, can be used, in principle,
to distinguish a black hole from an $\varepsilon$-wormhole which
mimics it. Suppose a small mass falling freely into a massive body,
such as an $\varepsilon$-wormhole.  Consider, for example, the
situation BT. One can compare two approaches. In the first approach,
one neglects the size of the small mass and considers the geodesic
along which such a point-like small mass moves \cite{ds}.  Then, if
the throat is very close to the would-be horizon and subsequent pulses
are emitted near the throat, the intervals of time measured at
infinity, grow unbound as $\Delta t\sim -\ln \varepsilon $ in the
limit $ \varepsilon \rightarrow 0$. So an observer at infinity cannot
distinguish the fall of matter into a wormhole with vanishingly small
$\varepsilon $ from absorption of matter by a black hole, if the
interval of observation time is less than $\Delta t$, by construction
a very long interval, see \cite{ds} for details.  In the second
approach, however, one takes the finiteness of the size of the
free-falling small mass into account. Then, the overall picture
changes since due to the growing tidal forces the small mass gets
deformed. If the small mass is a luminescent source, such a change can
in principle be detected by an observer at infinity. Moreover, such
changes happen much quicker than the typical times needed to penetrate
the immediate vicinity of a horizon. The intervals of time during
which the size of a small mass changes can be estimated from
Eq.~(\ref{kinf}) and the geodesic deviation equation. For a process
occurring near the throat it yields for the proper time the value
$\Delta \tau \sim r_{+}\varepsilon $, so that $\Delta \tau \rightarrow
0$, as $\varepsilon \rightarrow 0$. The corresponding interval of time
$ \Delta t$ at infinity is $\Delta t=\frac{\Delta \tau
}{\sqrt{-g_{00}}}\sim r_{+}$ is finite. Therefore, in case of
stretching, an observer at infinity would see an extended distorted
image of an infalling small mass. If the tidal forces are infinite, as
they can be in some of the cases shown above for
$\varepsilon$-wormholes, then the stretching is correspondingly
large. Of course, it is also possible that infinite contraction in
some direction converts an extended small mass to zero size in this
direction, see \cite{br1}. For the usual black holes, tidal forces are
also present but such forces are much weaker than for
$\varepsilon$-wormholes, where near the throat they can be as large as
one likes for sufficiently small $\varepsilon$. Thus, the key point is
to look not to single geodesics as in \cite{ds}, but to the separation
between geodesics of the same congruence. Such a separation delivers,
to an observer at infinity, meaningful information about the region of
strong gravity where a black hole or a $\varepsilon$-wormhole can be
situated, and such information has very different properties, which
depend on the massive body in question.

Second, we indicate what further physical
changes could be expected by performing some interesting estimates
of the effects. Here, one can take advantage of known results (see
\cite{MTW}, section 32.6). Suppose then that a box of small proper mass
$m$ and proper length $l$ is freelly falling towards an object of
mass $M$. To simplify we consider an uncharged object, $Q=0$.  From
\cite{MTW} one finds that the radial tidal force exerted on the box is
equal to $F=-\frac{1}{4}m lK$ where $K$ has the meaning of a tidal
radial acceleration. Indeed, $K$ is given precisely in
Eq. (\ref{K(r)}).  For the Schwarzschild metric one has
$K=\frac{2M}{r^{3}}$, so that near the horizon of a Schwarzschild
black hole, where $r\approx r_+=2M$, one obtains the value $K_{\rm
bh}\approx \frac{1}{4M^{2}}$. However, for Model 1, if we
correspondingly choose the metric potential as $V(r)=1-\frac{2M}{r}$ in
the situation BT, it follows from (\ref{kinf}) that near the would-be
horizon one has $K\approx\frac{1}{4\varepsilon ^{2}M^{2}}=
\frac{K_{\rm bh}}{\varepsilon^{2}}$.  Let us suppose that a distant
observer is able to recover from observational data the value of
$K$. If the observer thinks that the measured value of $K$ is due to a
Schwarzschild black hole, he should ascribe a mass $M_{\rm bh}$ to it.
However, if the object turns out to be a black hole mimicker, then,
for the same $K$, the value of the actual mass will be much greater,
given by $M_{\rm m}\sim \varepsilon ^{-1}M_{\rm bh}$.  If, in
addition, the observer, knowing the mass $M$ of the object, in this
case $M=M_{\rm m}$, insists in explaining it in terms of the usual
black hole metric, he will find that $M_{\rm bh}\ll M_{\rm m}$, and
will certainly start asking about the ``hidden mass". This example
shows that ``hidden mass" in some situations may arise simply because
the metrics of a black hole or of a black hole mimicker were not
properly discerned.

\subsection{Gravastars}

As far as gravastars are concerned, they contain, by construction, a thin
layer of normal matter with positive density $\rho $ and positive
radial pressure $p$ on the border of the tension matter with
vacuum. In the model suggested in \cite{grav} the stiff matter
equation of state with $p=\rho $ was chosen. Then, it follows
immediately from the field equations and the conservation laws that,
as the border approaches the gravitational radius, the gradient of the
pressure becomes infinite. One can, in general, admit discontinuous
radial pressure, giving rise to a surface pressure on the boundary
between matter and vacuum. However, this surface pressure and other
surface stresses also grow unbound in the horizon limit. Surely, the
more the border approaches the horizon, the better a black hole
mimicker it becomes, but, at the same time, the closer the system
approaches the singular state. This does not exclude in advance the
astrophysical significance of gravastars as compact vacuum-like
geometries but it shows that they can hardly pretend to be good black
hole mimickers.

In more detail, to see that the surface stresses go unbound, we
introduce the quantity $b=\exp(\Phi)$, so that we can rewrite
Eq. (\ref{22}) as $8\pi S_{}=\frac{1}{r_{0}}\left[\sqrt{V(r_{0}+0)}-
\sqrt{V(r_{0}-0)}\right]+ \frac {1 }{b(r_{0})}\left[\left(
\frac{db}{dl}\right) _{+}-\left( \frac{db}{ dl}\right) _{-}\right]$,
where we have taken into account that the continuity of the first
fundamental form demands $b_{+}=b_{-}=b(r_{0})$. In the outer region
we have the Schwarzschild metric, so $\left(\frac{db}{dl} \right)
_{+}>0$. Let the radius of the shell approach that of the would-be
horizon, i.e., $r_{0}\rightarrow r_{+}$, and $b(r_{+})=0$. By
definition of a gravastar, actually there is no horizon in the system,
so the function $b$ cannot cross $r$ at $r_{+}$ at all. In the inner
region, either $\left( \frac{db}{dl} \right) _{-}=0$ or $\left(
\frac{db}{dl}\right) _{-}<0$. Thus, since there is a $1/b$
$(r_{0}\rightarrow r_{+})$ term in $8\pi S_{}$, and the other terms
remain finite and non-zero, we find that $8\pi S\rightarrow \infty
$. This makes the gravastar unphysical in the near-horizon limit. It
is worth noting that this conclusion and its derivation is very close
to the statement that quasi-black holes cannot be non-extremal, if
only finite surface stresses are allowed (see Sec. IV of
\cite{qbh}). The only difference is that in the whole inner region
$\left( \frac{db}{dl}\right) _{-}\rightarrow 0 $ everywhere for
quasi-black holes whereas for gravastars $ \left( \frac{db }{ dl}
\right) _{-}$ can be non-zero there. However, in the present context,
only the vicinity of the would-be horizon is relevant, so the
conclusions are similar.

\section{Mimickers of extremal black holes}

\subsection{Wormholes on the basis of extremal $\protect\varepsilon$-metrics
with surgery}

\subsubsection{Basics}

\noindent In Section \ref{nonextremalepsilonsurg} on wormholes on the basis
of nonextremal $\varepsilon$-metrics with surgery, we have discussed wormhole
configurations which mimic non-extremal black holes. Here we consider the
one-parametric deformation (\ref{1}) such that for $\varepsilon=0$ our
metric represents the gluing of two extremal Reissner-Nordst\"{o}m black
holes. Note that in the non-extremal case, when using metric (\ref{1}) we
have chosen a function $V$ which itself does not contain the parameter $
\varepsilon$. However, now such a simple construction cannot be implemented.
Our goal is to trace the relationship between a wormhole metric with a
generic $\varepsilon$ parameter ($\varepsilon\neq 0$), and a black hole
metric ($\varepsilon=0$). If we start from equation (\ref{1}) with $
r_{+}=r_{-}$, i.e., $V=(1-\frac{ r_{+}}{r})^{2}$, we encounter the immediate
difficulty that spacetime is geodesically complete and represents an
infinitely long horn, so one cannot speak about a wormhole at all.
Therefore, we should deform the extremal Reissner-Nordstr\"{o}m metric in a
somewhat different way and include the parameter $\varepsilon$ not only into
the function $\Phi $ but into $V$ as well. Let us make the simplest choice
for $V(r,\varepsilon)$, namely,
\begin{equation}
V(r)=\left( 1-\frac{r_{+}}{r}\right) \left( 1-\frac{r_{-}}{r}\right)
\,,\quad \mathrm{with}\quad r_{-}=r_{+}\left( 1-\delta (\varepsilon)\right)
\,,  \label{vne}
\end{equation}
$\delta (\varepsilon)$ being such that $0\leq\delta(\varepsilon)\leq 1$, and
$\delta (0)=0$. At some $r_{0}>r_{+}$, we glue two copies of the spacetime,
the ``$+$''\ branch to the ``$-$'' branch. Then, the behavior of $K$ and $S$
follows from Eqs. (\ref{e}) and (\ref{22e}). Note that if $\delta =
\mathrm{
constant}$ and $\delta \leq 1$ we return to the deformed non-extremal
Reissner-Nordstr\"{o}m case and the results (\ref{einf})-(\ref{0}) are
reproduced, $\delta =1$ being the Schwarzschild case.

\subsubsection{Models}

\vskip 0.3cm \hskip 7.4cm \textit{\small Model 1} \vskip 0.1cm

\noindent Using (\ref{efv}), i.e., $\exp(\Phi)=\sqrt{V+\varepsilon^{2}}$,
and (\ref{vne}), we can study Model 1 in the extremal case. \vskip0.4cm
\noindent \textit{Situation BT:} As for the general behavior one finds,
\begin{equation}
\lim_{\varepsilon\rightarrow 0}\lim_{r_{0}\rightarrow
r_{+}}K(r_{0},\varepsilon)=-\frac{\alpha }{4r_{+}^{2}}\,,\quad \mathrm{with}
\quad \alpha =\lim_{\varepsilon\rightarrow 0}\left( \frac{\delta }{
\varepsilon}\right) ^{2}\,,  \label{eqK}
\end{equation}
and,
\begin{equation}
\lim_{\varepsilon\rightarrow 0}\lim_{r_{0}\rightarrow r_{+}}8\pi
S_{{}}(r_{0},\varepsilon)=0\,.
\end{equation}
Let us analyze the naked behavior. In principle, the quantity $\alpha
$ may be finite or infinite depending on the model for $\delta
(\varepsilon)$. From Eq. (\ref{eqK}), longitudinal contraction is
finite if $\alpha $ is finite, or infinite if $\alpha $ is
infinite. The value of $\bar{N}$ which determines transverse
deformation can be found from eq. (\ref{newbarN}).  Then, we obtain
that in the limit under discussion
\begin{equation}
\bar{N}=\frac{\beta E^{2}}{2r_{+}^2}\,,  \label{nb}
\end{equation}
where $\beta =\lim_{\varepsilon\rightarrow 0}\frac{\delta
}{\varepsilon ^{2}}$. Then we have transverse stretching which is
finite if $\beta$ is finite, or infinite if $\beta $ is infinite. The
stress component $\bar\Sigma$ is finite, indeed zero, in the limit.

\vskip0.4cm \noindent \textit{Situation TB:} As for the general behavior one
finds,
\begin{equation}
\lim_{r_{0}\rightarrow r_{+}}\lim_{\varepsilon \rightarrow
0}K(r_{0},\varepsilon )=-\frac{1}{r_{+}^{2}}\,,
\end{equation}
\begin{equation}
\lim_{r_{0}\rightarrow r_{+}}\lim_{\varepsilon \rightarrow
0}S_{{}}(r_{0},\varepsilon )=\frac{1}{4\pi r_{+}}\,.  \label{2bt1}
\end{equation}
Note that when $\alpha =4$ in the situation BT, the quantities $K$ for two
situations coincide. Let us analyze the naked behavior. Here there is finite
longitudinal contraction, and no transverse deformation since, according to
eq. (\ref{newbarN}), $\bar{N}\rightarrow 0$ in the limit under
consideration. Thus, the only manifestation of naked behavior is connected
with the surface stresses. According to (\ref{sigmacapital2}), the quantity
$
\bar{\Sigma}$ behaves as
\begin{equation}
\bar{\Sigma}\rightarrow -\infty \,,
\end{equation}
i.e., it diverges in situation TB.

\noindent \vskip0.3cm \hskip7.4cm \textit{\small Model 2} \vskip0.1cm
\noindent Using Eq. (\ref{Phimodel2}), i.e., $\exp(\Phi)=\lambda\sqrt{V}
+\varepsilon$, and (\ref{vne}), we can study Model 2 in the extremal case.

\noindent \textit{Situation BT:} As for the general behavior one finds,
\begin{equation}
\lim_{\varepsilon \rightarrow 0}\lim_{r_{0}\rightarrow
r_{+}}K(r_{0},\varepsilon )=0\,,
\end{equation}
\begin{equation}
\lim_{\varepsilon \rightarrow 0}\lim_{r_{0}\rightarrow
r_{+}}S_{{}}(r_{0},\varepsilon )=\frac{\lambda \sqrt{\alpha }}{4\pi r_{+}}
\text{ .}
\end{equation}
Let us analyze the naked behavior. There is no deformation in the radial
direction. The behavior of $\bar{N}$ which is responsible for transverse
deformation, can be obtained from (\ref{n2}) and coincides with (\ref{nb})
in the limit under discussion. Thus, again transverse stretching is finite
if $\beta $ is finite, or infinite if $\beta $ is infinite. Note that it
follows from the definitions of $\alpha $ and $\beta $ that $\alpha
=\lim_{\varepsilon \rightarrow 0}\beta \delta $. According to the definition
(\ref{vne}) of $\delta (\varepsilon )$, $\lim_{\varepsilon \rightarrow
0}\delta (\varepsilon )=0$. Therefore, if $\beta $ is finite, $\alpha =0$,
then $K$, $S$, $\bar{N}$ and $\bar{\Sigma}$ are finite (moreover, $S=\bar{
\Sigma}=0$), so naked behavior is absent in this case. On the other hand, as
$\beta =\lim_{\varepsilon \rightarrow 0}\frac{\alpha }{\delta }$, in case $
\alpha \neq 0$ the quantity $\beta $ diverges and so does $\bar{N}$. This
means transverse stretching is infinite.

\vskip0.4cm \noindent \textit{Situation TB:} As for the general behavior one
finds,
\begin{equation}
\lim_{r_{0}\rightarrow r_{+}}\lim_{\varepsilon \rightarrow
0}K(r_{0},\varepsilon )=-\frac{1}{r_{+}^{2}}\,,
\end{equation}
\begin{equation}
\lim_{r_{0}\rightarrow r_{+}}\lim_{\varepsilon \rightarrow
0}S_{{}}(r_{0},\varepsilon )=\frac{1}{4\pi r_{+}}\,.  \label{2bt2}
\end{equation}
Let us analyze the naked behavior. One finds, finite contraction in the
longitudinal direction and no transverse deformation. However,
\begin{equation}
\bar{\Sigma}\rightarrow -\infty \,,
\end{equation}
i.e., it diverges.

\vskip 0.5cm \centerline{\it\small Overall conclusions for Models 1 and
2}

\vskip0.1cm \noindent As an overall conclusion, assuming that for
$\varepsilon =0$, one has $\exp (2\Phi)=V,$ and taking into account
that $ V(r_{+})=0$, $\exp [\Phi (r_{+},\varepsilon )]\neq 0$, we
obtain immediately from (\ref{sigmacapital2}) that in the situation
BT, $\bar{\Sigma}=0$, and in the situation TB, $\bar{\Sigma}$
diverges. This holds independently of the kind of the model, and is
valid for the non-extremal spacetimes, as well as for the extremal
spacetimes. Thus, a free-falling observer encounters diverging surface
energy density in the situation TB. The same conclusion applies to the
flux $J=S_{\mu \nu }\bar{u}^{\mu }\bar{n}^{\nu }$.


\subsubsection{Remarks: Light-like shells and classical electron models}

\vskip 0.1cm \centerline{\it\small Light-like shells} \vskip 0.1cm

\noindent One essential feature of the almost extremal configurations
under discussion consists in that we deal with shells which are
timelike but become light-like in the limit when they are approaching
the would-be horizon. A natural question arises: what happens if we
start out our analysis with two extremal black hole spacetimes, i.e.,
putting $ \varepsilon=0$ and $r_0=r_{+}$, and the shell in-between the
two spacetimes lies on the event horizon and so is lightlike? More
precisely, we are interested whether the stresses on the shell remain
finite or become infinite. One can expect from our previous results
that they are finite but this is not so obvious in advance. We have
seen already that taking different limiting procedures in the
near-horizon limit is a rather delicate issue.  Moreover, the
formalism for lightlike shells \cite{pois,bar} somewhat differs from
that for timelike ones, so we perform our analysis anew. We restrict
ourselves to the case of taking two extremal Reissner-Nordstr\"{o}m
black holes. The reason for this choice is that only extremal black
holes are good candidates for gluing without severe singularities,
i.e., the gluing procedure maintains a finite Kretschmann scalar
throughout the spacetime as well as finite stresses on the glued
surface. This case is also a Majumdar-Papapetrou case, and could be
analyzed in the next section as well, but since it is also a limit of
what has been done above, namely the null limit of a timelike wormhole
throat with two extremal external vacuum spacetimes, we discuss it
now. To match two extremal Reissner-Nordstr\"{o}m spacetimes at the
light-like surface $r_0=r_{+}$ we follow the general formalism for
light-like shells \cite{pois,bar}. We write the metric in Kruskal-type
coordinates $ds^{2}=-H(U,V)\,\,dUdV+r^{2}(U,V)\,\,(d\theta ^{2}+\sin
^{2}\theta d\phi ^{2}).$ Let the surface $r_0=r_{+}$ correspond, say,
to $U=0$. Then, the effective energy density is $\mu =-\frac{\gamma
_{\theta \theta } }{16\pi r^{2}}$, (see, e.g., equation (3.99) of
\cite{pois}). Here $\gamma _{\theta \theta }=[\left( \frac{\partial
g_{\theta \theta }}{ \partial x^{\alpha }} \right) _{+}-\left(
\frac{\partial g_{\theta \theta }}{ \partial x^{\alpha }} \right)
_{-}]N^{\alpha }$, where the indexes $``+"$ and $``-"$ refer to the
different sides of the shell, and the null vector $ N^{\alpha }$ is
such that $k_{\alpha }N^{\alpha }\neq 0$, with $k^{\alpha }$ being a
null tangent vector. Now, the only nonvanishing components of $
k^{\alpha }$ and $N^{\alpha }$, are by construction $k^{V}$ and
$N^{U}$.  Since there is no rest frame in the null case, the measured
energy density $ \mu $ depends on the chosen observer. To check that
it is finite and non-zero, it is sufficient to check that $[\left(
\frac{\partial r}{\partial U}\right) _{+}-\left( \frac{\partial
r}{\partial U}\right) _{\_}]$ is finite and non-zero. For this
purpose, it is sufficient to exploit the result of \cite{vf} where it
is shown that $\frac{\partial r}{\partial U}=-1$ on the horizon. In
our case, as we deal with two black holes instead of a single one, the
coordinate $r$ increases on both sides of the shell. Thus, $\frac{
\partial r}{\partial U}$ has different signs and takes the values $\pm
1$ on each side. So indeed, the difference is equal to $2$ and is
finite and non-zero. In general, there are other contributions to the
effective stress-energy tensor of the shell due to effective pressures
and currents, but it is easy to show that in our case they are
absent. It is worth noting that the gluing in our case differs from
the gluing between different extremal Reissner-Nordstr\"{o}m black
holes considered in \cite{dray}, in that we replaced the usual metric
inside the horizon by their $``-"$ branch.  It also differs from the
wormhole construction used in \cite{al} where two spheres were cut out
from a vacuum Majumdar-Papapetrou system, or more precisely, they were
cut from a single spacetime containing two extremal
Reissner-Nordstr\"{o}m black holes.

\vskip 0.3cm \centerline{\it\small Classical electron models} \vskip 0.1cm

\noindent As a by-product of this light-like shell construction, and a
very interesting one, we have just found a configuration that
represents a regular wormhole configuration which is also a black
hole. More important perhaps, in addition, it can serve as a classical
model for an elementary particle in that (i) the system is
characterized by a minimum number of fixed parameters like mass and
charge and (ii) it is free of a central singularity inside. We are
aware that it is not entirely of electromagnetic nature because it has
stresses on the horizon, a kind of Poincar\'{e} stresses. But, anyway,
such a surface stress, can be considered as a mild singularity when
compared to the usual central singularity. In this model a
free-falling observer can penetrate to the inside but, of course,
cannot return back to the original asymptotic region, due to the
existence of a horizon. So the wormhole is an untraversable one. Thus,
the body under discussion combines features of an untraversable
wormhole and a regular black hole, and can be called a worm-black
hole. Since the proper distant to the extremal horizon is infinite,
such a hybrid construction is similar to the null wormholes, or
N-wormholes for short, see \cite{n}. Note that it differs from
configurations representing quasi-black holes. For example, in some
quasi-black hole models, see \cite{qbh}, there is a region $r\leq
r_{+}$ which becomes degenerate in the quasihorizon limit. This region
is missing in our worm-black hole model, since the black hole metric
beyond the horizon is replaced by a branch with an areal radius that
grows away from the horizon.  It also differs from the model
considered in \cite{vf} where the external Reissner-Nordstr\"{o}m part
was glued to the Bertotti-Robinson metric that leads to surface
stresses that vanish in the horizon limit in the static frame but grow
unbound in the free-falling one, and as a result, the inner region
becomes impenetrable for a free-falling observer (see \cite{qbh} for
details).

\subsection{Quasi-black holes}

Another candidate for the role of a black hole mimicker is a
quasi-black hole. Roughly speaking, it is an extremal object that
appears when the system approaches the quasihorizon as nearly as one
likes, along a family of quasi-static configuration. It was argued in
\cite{qbh} that such a limit can correspond to an extremal
quasihorizon only if we restrict ourselves to static configurations
which are regular in the strong sense. The latter means that the
Kretschmann scalar should be finite everywhere in the system, and
surface stresses at the quasihorizon should be finite as well. There
are subtleties in the non-trivial relation between regular and
singular features of quasi-black holes \cite{qbh}. For example, the
whole region can look degenerate from the viewpoint of a distant
observer and, nevertheless, the Kretschmann scalar remains finite in
that region.

In more detail, consider the static spherically symmetric metric
(\ref{1}), and let it represent a spacetime in which there is an inner
matter configuration, attached to an asymptotic flat exterior
region. The $\lambda _{i}$ in (\ref{1}) stand for the radius $r_{0}$
of the configurations and possibly some other parameters connected
with the particular object one is analyzing. For instance, the
parameter $\varepsilon$ also enters in the analysis, and here has a
slightly altered meaning. It means a small deviation from a
quasi-black hole, rather than from a black hole solution as in section
\ref{nonextremalepsilonsurg}, see also below. Suppose the spacetime in
question has the following properties: (a) the function $V(r)$ in
(\ref{1}) attains a minimum at some $r^{\ast }\neq 0$, such that
$V(r^{\ast })=\varepsilon$, with $\varepsilon<<1$, this minimum being
achieved either from both sides of $r^{\ast }$ or from $r>r^{\ast }$
alone, (b) for such a small but nonzero $\varepsilon$ the
configuration is regular everywhere with a nonvanishing metric
function $\exp (2\Phi )$, at most the metric contains only
delta-function like shells, and (c) in the limit $\varepsilon
\rightarrow 0$ the metric coefficient $\exp (2\Phi )\rightarrow 0$ for
all $ r\leq r^{\ast }$. These three features define a quasi-black
hole. In turn, these three features imply that, there are infinite
redshift whole regions when $\varepsilon\rightarrow 0$, a free-falling
observer finds in his own frame infinitely large tidal forces in the
whole inner region, showing thus naked behavior, although the
curvature scalars are finite. Moreover it has some form of degeneracy
since, although the spacetime curvature invariants remain perfectly
regular everywhere, in the limit, outer and inner regions become
mutually impenetrable and disjoint. For a free-falling external nearby
observer it is as if a null singular horizon is being formed. For
external far away observers the spacetime may be said to be naively
indistinguishable from that of extremal black holes. However, if one
makes experiments with infalling luminescent extended small masses,
one might find differences, since as discussed previously \cite{qbh},
due to the naked behavior, quasi-black holes enlarge grossly the tidal
forces on an infalling small mass when compared to the tiny effect of
an extremal black hole on the same small mass. Thus, as with the
extremal $\varepsilon$-metrics studied before, the naked behavior
shows that quasi-black holes are not so good mimickers, as was
previously thought, but they are still better than black foils
\cite{ds} where the singularity is more severe. A further important
property is that quasi-black holes must be extremal.  For a
quasi-black hole the metric is well defined and everywhere
regular. However, when $\varepsilon=0$, quasi-black hole spacetimes
become degenerate, almost singular, see \cite{qbh}.  The quasi-black
hole is on the verge of forming an event horizon, but it never forms
one, instead, a quasihorizon appears.  In summary quasi-black holes
have normal general behavior and singular naked behavior.  Quasi-black
holes may appear from Bonnor stars, i.e., systems composed of extremal
charged dust and vacuum, from self-gravitating Higgs magnetic monopole
systems, and from composite spacetimes even in the case of pure
electrovacuum, in which these vacuum systems are composed of an
exterior Reissner-Nordstr\"{o}m part glued to an inner
Bertotti-Robinson spacetime or of an exterior Reissner-Nordstr\"{o}m
part glued to an an inner Minkowski spacetime, see \cite{qbh} for a
full discussion and references.

\subsection{Wormholes on the basis of quasi-black holes from Bonnor stars}

\subsubsection{Basics}

\noindent Bonnor stars are Majumdar-Papapetrou matter systems with either a
sharp or smooth boundary to an exterior vacuum. Since Bonnor stars are
paradigmatic to understand the formation of quasi-black holes (see
\cite{qbh}), 
it is interesting to use those stars on the threshold of forming a
quasi-black hole to understand whether wormholes on the basis of quasi-black
holes which can be formed from Bonnor stars, can mimic extremal black holes
or not. This is an interesting variant, although with similarities, to
wormholes on the basis of extremal $\varepsilon$-metrics. We use Bonnor
stars, both in their compact version \cite{bonnor,lemoszanchind}, as well as
in their extended one \cite{extdust}.

\subsubsection{Wormholes on the basis of quasi-black holes from compact
Bonnor stars}

\noindent Now, we start from the configuration which contains
Majumdar-Papapetrou matter inside and vacuum outside, a compact Bonnor
star \cite{bonnor,lemoszanchind}. Let us have a compact object, a
Bonnor star, with extremal dust for $r\leq r_{0}$, joined to an
extremal Reissner-Nordstr\"{o}m metric for $r\geq r_{0}$.  The
potential $V(r) $ can be written as
\begin{equation}
V(r)=\left( 1-\frac{\mu (r)}{r}\right) ^{2}\,,  \label{vm}
\end{equation}
with the mass density $\rho$ and the function $\mu(r)$ being connected
through $4\pi \rho =\frac{\mu ^{\prime }}{r^{2}}\left( 1-\frac{\mu
}{r} \right)$. The function $\mu(r)$ can be interpreted as the proper
mass enclosed within a sphere of a radius $r$. In addition, for
extremal dust, one has $\mu(r)=e(r)$, where $e(r)$ is the electric
charge within this sphere. At the boundary $r_0$, one has $\mu
(r_{0})=M$, and $e(r_{0})=Q$, such that $M=Q$, $M$ and $Q$ being the
total mass and charge, respectively.

To construct a wormhole, one can take the following procedure. Cut the
solution somewhere in the interior at some radius $r_{1}<r_{0}$ and
discard the region $r<r_{1}$. One obtains a Majumdar-Papapetrou matter
region for $ r_{1}\leq r<r_{0}$, and a vacuum region for $r_{0}\leq
r<\infty $. For definiteness, let this spacetime be situated on the
left. A symmetric right branch, also containing interior and exterior,
is again cut at the radius $ r_{1}<r_{0}$, and glued to the symmetric
left branch, producing thus a boundary shell at $r_{1}$. Then, it
follows from the field equations that the left branch is given by
\begin{eqnarray}
\exp (\Phi ) &=&\exp \left[ \int_{r_{0}}^{r}dr\,\frac{\mu }{r^{2}\,
\left( 1-
\frac{\mu }{r}\right) }\right] \,,\quad r_{1}\leq r\leq r_{0}\,,  
\notag \\
\exp (\Phi ) &=&1-\frac{M}{r}\,,\quad r_{0}\leq r<\infty \,,  
\label{Phibon}
\end{eqnarray}
and that the proper radius $l$ and the coordinate radius $r$ are 
related by,
\begin{equation}
\frac{dr}{dl}=-\left( 1-\frac{\mu }{r}\right) \,.  \label{r+-1}
\end{equation}
As the Bonnor star is here a compact object, the proper distance from
$r_{1}$
to $r_{0}$ is finite, and tends to zero in the limit $r_{1}\rightarrow
r_{0}$. 
Therefore, in this limit, the matter between right and left boundaries
becomes negligible and the construction corresponds to gluing two
extremal Reissner-Nordstr\"{o}m black holes in the situation TB of
section \ref {nonextremalepsilonsurg}. Thus in the limit of our
interest, $r_{1}\rightarrow r_{0}\rightarrow M$, it is not surprising
that the results coincide with (\ref{2bt1}) and (\ref{2bt2}) where
$r_{+}=M$. Indeed, one finds
\begin{equation}
K=\mathrm{finite}\,,  \label{k-1}
\end{equation}
and
\begin{equation}
8\pi S_{{}}=\frac{2}{Q}=\frac{2}{M}\,,  \label{+-2-1}
\end{equation}
being finite as well. We have put $M=Q$, as is the case 
for these systems.

There is yet another procedure to produce a wormhole. In the above
considerations, we performed a symmetric construction, in the sense that the
$``+"$ and $``-"$ branches differed by the sign of $\frac{dr}{dl}$ only.
Now, we start again from a compact Bonnor star configuration which contains
matter inside and vacuum outside. We want to preserve this feature, and so
we have to make a non-symmetric deformation. Thus, we change the procedure
and consider the following construction. Again, a Bonnor star is made of
Majumdar-Papapetrou matter for $r\leq r_{0}$, which in turn is joined to an
extremal Reissner-Nordstr\"{o}m metric for $r\geq r_{0}$. Now, in the region
$r\leq r_{0}$, the left $``-"$ branch, choose the distribution with $\frac{
dr }{dl}\leq 0$. The potential $V(r)$ can be written again as in Eq. (\ref
{vm}). Then, for the left branch, it follows from the field equations that
\begin{equation}
\exp (\Phi )=\exp \left[ \int_{r_{0}}^{r}dr\,\frac{\mu }{r^{2}\,\left( 1-
\frac{\mu }{r}\right) }\right] \,,\quad 0\leq r\leq r_{0}\,,
\label{Phibon2}
\end{equation}
and
\begin{equation}
\frac{dr}{dl}=-\left( 1-\frac{\mu }{r}\right) \,.  \label{r+}
\end{equation}
It is worth paying attention that because of the property
$\frac{dr}{dl}<0$, the matter distribution which was originally
compact, turned after deformation into a non-compact one since at left
infinity $l\rightarrow -\infty $, where \ $\mu \rightarrow 0$,
$\frac{dr}{dl}\rightarrow -1$. For the right $``+"$ branch we use the
extremal Reissner-Nordstr\"{o}m metric with the mass $M=Q$, which
gives,
\begin{equation}
\exp (\Phi )=1-\frac{M}{r}\,,\quad r_{0}\leq r<\infty \,,  \label{Phibon3}
\end{equation}
and
\begin{equation}
\frac{dr}{dl}=1-\frac{M}{r}\,.  \label{r-right}
\end{equation}
In the limit $r_{0}\rightarrow M$, $\mu \rightarrow M=Q$ one finds that
\begin{equation}
K=\mathrm{finite}\,,  \label{k}
\end{equation}
and
\begin{equation}
8\pi S_{{}}=\frac{2}{Q}=\frac{2}{M}\,,  \label{+-2}
\end{equation}
being finite as well. We have put $M=Q$, as is the case 
for these systems.

Thus, extremal quasi-black wormholes, made of Majumdar-Papapetrou matter are
possible. Their distinctive feature is the presence of finite non-zero
surface stresses on the horizon. Curvature components remain finite. It is
worth noting that, although in this subsection we did not introduce the
parameter $\varepsilon$ explicitly, actually its role is played, say, by the
difference $r_{0}-M$ in the sense that this quantity is responsible for the
deviation of the spacetime from its limiting state (a quasi-black hole or
two quasi-black holes glued together).

\subsubsection{Wormholes on the basis of quasi-black holes from extended
Bonnor stars}

\noindent 
Here we exploit the distribution of extremal charged dust given in \cite
{extdust}. Near the quasi-black hole limit, the first order corrected
quasihorizon has radius $r^{\ast }$ given by
\begin{equation}
r^{\ast }=q\left[ 1+\frac{3}{4}\left( \frac{2c^{2}}{q^{2}}\right) ^{1/3}+...
\right] \,,  \label{radiusquasi}
\end{equation}
where $c$ is the parameter that yields the deviation from the
Reissner-Nordstr\"{o}m solution, and $q$ is a quantity with units of
electric charge, which is indeed the total charge $Q$ when $c=0$ from the
outset (see \cite{qbh,extdust} for details). The quasi-black hole limit is
such that $c<<q$, with $c\rightarrow 0$. In a sense, the dimensionless
parameter $c/q$ here corresponds to the $\varepsilon$ parameter in the $
\varepsilon$-wormhole construction of a previous section. Then the solution
has asymptotics near the quasihorizon $r^{\ast }$ given by
\begin{equation}
V=\frac{9}{2^{4/3}}\left( \frac{c}{q}\right) ^{4/3}+\frac{2(r-r_{\ast })}{
q^{2}}^{2}+...\,,  \label{as1}
\end{equation}
and
\begin{equation}
\exp (\Phi )=2^{1/3}\left( \frac{c}{q}\right) ^{2/3}+\frac{2}{3}\frac{
(r-r_{\ast })}{q}+\frac{2^{2/3}}{9c^{2/3}q^{4/3}}\left( r-r_{\ast }\right)
^{2}...\,.  \label{as2}
\end{equation}
Consider again the $``+"$ and $``-"$ branches, symmetric relative to the
first order corrected quasihorizon radius in the region $r\geq r^{\ast }$,
but with different signs of $\frac{dr}{dl}$. Each branch has the same
dependence of the metric potential, Eqs. (\ref{as1})-(\ref{as2}), on $r$.
Thus, for simplicity, we restrict ourselves to the analog of the situation
BT considered previously, which translated to here means taking the limits
as follows, $\lim_{\frac{c}{q}\rightarrow 0}\lim_{r_{\ast }\rightarrow
r_{+}} $. Then, simple calculations show that
\begin{equation}
K=\mathrm{finite}\,,
\end{equation}
and
\begin{equation}
8\pi S_{{}}=\frac{2}{Q}=\frac{2}{M}\,,  \label{lim}
\end{equation}
being finite as well. In (\ref{lim}) we have put $Q=q$ in the limit $
c\rightarrow 0$ and $M=Q$ also in this limit, where $M$ is the mass of the
configuration. In the limit under consideration the metric in the region $
r>r^{\ast }$ is given by the extremal Reissner-Nordstr\"{o}m metric, whereas
in the immediate vicinity of the quasihorizon the metric is described by the
Bertotti-Robinson metric. Therefore, as in the preceding subsection, our
construction gives two extremal Reissner-Nordstr\"{o}m black holes glued
along the quasihorizon with different signs of $\frac{dr}{dl}$ on opposite
sides.

\section{Discussion and conclusions}

We have studied wormhole and other configurations as possible
mimickers of black holes. We have separated the configurations into
non-extremal and extremal.

For wormholes, we have examined separately two limiting procedures in
which the wormhole throat approaches the black hole horizon. In the
first procedure, we fix the location of an observer (exactly on the
throat). Then, we change the spacetime (making a wormhole on the verge
of being a black hole). This is the situation BT. In the second, we
change spacetime (making a wormhole on the verge of a black hole),
then place the shell outside the throat and move it toward the throat
(which coincides with the horizon).  This is the situation TB. This
procedure is carried out for non-extremal and extremal configurations
separately. In the non-extremal case it turned out that each of the
limits under discussion is singular: either the Kretschmann scalar or
surface stresses on the throat (or both) diverge. Thus, the limit is
singular. In other words, a mimicker of a non-extremal black hole,
made from a non-extremal wormhole, including the black foil of
\cite{ds}, is not smooth. We have summarized the results for the
non-extremal case in a table.  For the extremal case both the
Kretschmann scalar and surface stresses remain finite. This pronounced
distinction between properties of the limiting configurations in the
non-extremal and extremal cases is one counterpart of the conclusion
made in \cite{qbh} that quasi-black holes can be only
extremal. However, one should not forget about some subtleties
connected with the fact that singular behavior can, in general,
manifest itself not only in the value of the Kretschmann scalar. Even
if this scalar is finite, a naked behavior is possible or even
inevitable as was shown in in the present paper and in \cite{qbh}
(see, e.g., Sec. V of \cite{qbh} for a discussion about other
subtleties in which the singular features of quasi-black holes 
are revealed). There is also another candidate for the role of a
non-extremal black hole mimicker, a gravastar \cite{grav}.  However,
the corresponding surface stresses grow unbound when the radius
approaches the gravitational radius, as we have seen.

{} From an astrophysical viewpoint, the situation BT in the
non-extremal case, i.e., a black foil, is more interesting since it
implies no necessity of making a shell by hand. It is the case
considered in \cite{ds}. One of the questions raised in \cite{ds} is
whether it is possible or not to distinguish between a true
non-extremal black hole, a Schwarzschild black hole say, and a
wormhole. The main conclusion of \cite{ds} is that it is impossible to
distinguish for any finite time in the limit under discussion. This
conclusion is reached on the basis of considering properties of bodies
moving along separate fixed geodesics and emitting signals detected at
infinity. However, if from single geodesics we shift our attention to
a congruence of geodesics, it turns out that the strong gravity
forces, on the near horizon region, leave their imprint on the form of
a moving body and, thus, on the properties of signals which an
observer at infinity is detecting. If the surface of the body is
luminescent, an observer at infinity would see either a finite width
instead of a point, a continuous detection instead of separate pulses,
and so on. It is essential that in the case discussed in \cite{ds} the
corresponding proper time of deformation tends to zero when the
curvature grows unbound, with the time at infinity being finite. Thus,
at least in principle, an observer can distinguish between a black
hole and an almost singular wormhole. The singular nature of the limit
in the non-extremal case makes also questionable the applicability of
the membrane paradigm used in \cite {ds}. The key point in this
paradigm consists in boundary conditions according to which a
free-falling observer sees a finite value of physical fields on the
horizon (see \cite{th}, Sec. II). However, in the problem under
discussion, typically, this observer as well as the geometry itself
become ill-defined. Only in some cases (see Model 2, situation TB) the
curvature components remain finite in the free-falling frame. But even
in such situations the infinite surface stresses on the horizon
surface makes the physical meaning of the membrane paradigm unclear
since this paradigm relies heavily on the concept of a regular
surface.

Objects based on nearly extremal wormholes, although of less interest
astrophysically perhaps, have a much better behavior in the sense that
both the geometry and surface stresses remain finite. Moreover,
typically, there is no naked behavior. In this case, the effect of
strong curvature is much less pronounced than in the case of
quasi-black holes where a naked behavior is typical \cite{qbh}. In
this sense, a wormhole composed on the basis of two extremal black
holes seems to be the best mimicker of an extremal black hole. As
by-product, we have obtained a model of a regular black hole!

Thus, if we try to arrange a ranking of black hole mimickers, both
non-extremal and extremal, the list looks as follows from top to
bottom: wormholes on the basis of extremal black holes or on the basis
of quasi-black holes, quasi-black holes, wormholes on the basis of
non-extremal black holes (and within these the best are black foils),
and gravastars.  Bearing in mind that in observational astrophysics it
is difficult to find extremal configurations (the would be best
mimickers), whereas non-extremal configurations are really bad
mimickers, the task of distinguishing black holes from their mimickers
seems to be less difficult than one could think of it.

In the present paper we have restricted ourselves to static
spherically symmetric spacetimes. Meanwhile, in a recent work
\cite{cpcce} the status of black hole mimickers is undermined in the
rapidly rotating case as well since it is argued that they are
unstable. We have also circumscribed our discussion to particular
wormholes, gravastars and quasi-black holes
\cite{ds}-\cite{lemoszanchinjmp}, since these objects are well adapted
to our goal of examining the near-horizon properties and their
connection with far away asymptotic properties. However, there are
many other objects with properties that make them also potential black
hole mimickers (see, e.g., \cite{dars,guz,lib,sken}) and which are
worthy of study within our formalism.

\begin{acknowledgments}
O. Z. thanks Centro Multidisciplinar de Astrof\'{\i}sica--CENTRA for
hospitality and a stimulating working atmosphere. This work was partially
funded by Funda\c c\~ao para a Ci\^encia e Tecnologia (FCT) - Portugal,
through project POCI/FP/63943/2005.
\end{acknowledgments}

\end{document}